# Hyperspherical Approach to the Four-Body Problem

Nirav P. Mehta[*], Seth T. Rittenhouse, José P. D'Incao and Chris H. Greene

Dept. of Physics and JILA, University of Colorado, Boulder CO, USA
[*]E-mail: mehtan@colorado.edu

**Abstract:** The four-particle system is the simplest few-body system containing the fundamental physics involved in ultracold fermionic gases. We have made recent efforts to solve the quantum four-body problem in the adiabatic hyperspherical representation. Our approach yields a set of coupled potential curves that can in turn be solved for all elastic and inelastic processes. These rates may play an important role in the lifetime of molecules in ultracold fermi gases. We believe this will lead to a deeper understanding of ultracold fermi systems and the physics of the BCS-BEC crossover.

I. INTRODUCTION

The past few years have marked great experimental advances in the manipulation of degenerate Fermi gases. Several groups have succeeded in the formation [1-2] and condensation of [3-5] long-lived bosonic molecules in a Fermi gas of atoms  All such experiments make use of Feshbach resonances to tune the atom-atom scattering length a from the Bardeen-Cooper-Schrieffer (BCS) regime involving weak interactions between fermionic atoms, to the Bose-Einstein Condensation (BEC) regime of weakly interacting deeply bound bosonic molecules. During the passage from one regime to the other, the scattering length is taken through a so-called "crossover" regime where $|a| \to \infty$. It has been proposed that the observation of superfluidity in this regime would provide a connection between bosonic and fermionic superfluidity [6-7].  One quantity crucial to producing the correct equation of state in the bosonic limit, which many-body theories have failed to reproduce, is the value of the dimer-dimer scattering length $a_{DD}$.  Petrov et al [8] have solved for this quantity by correctly imposing the zero-range boundary conditions between interacting fermions, producing $a_{DD} \approx 0.6a$.

Our goal here is to understand the energy landscape of the four-body problem on the positive scattering length side of the Feshbach resonance. The four-body system is the simplest system involving two interacting dimers comprised of fermionic atoms. In order to simplify our calculations, we consider the following single channel two-body interaction that admits an analytic solution:

$$V(r) = -D/\cosh^2(r/L) \tag{1}$$

We perform a first-principles calculation of the adiabatic hyperspherical potential curves, calculate the dimer-dimer scattering length, and compare our result to that of Ref. [8].The four-body problem in 3 dimensions involves 12 total degrees of freedom, 3 dimensions for each particle. Jacobi coordinates described in Section III describe the various relative coordinates used to remove the center-of-mass motion. We shall further consider only states with total spherical symmetry, and hence need to isolate the dependence on the three Euler angles. This is accomplished by the introduction of body-fixed coordinates [9, 10] described in Section IV. We begin in section II by describing Jacobi coordinates for the four-body system, which allows us to remove the center of mass motion. These coordinates allow us to reduce the original 12-dimensional partial differential equation (PDE) to a 6-dimensional PDE. By then treating the hyperradius, R, as an adiabatic parameter (see Section II), it is possible to convert the 6-dimensional problem to a set of coupled equations in R alone.

## II. ADIABATIC HYPERSPHERICAL REPRESENTATION

The adiabatic hyperspherical representation [11] is similar in spirit to the Born-Oppenheimer representation in molecular physics. In this case, rather than the nuclear separation, the hyperradius R serves as an adiabatic parameter. The hyperradius is a quantity that describes the general size of the few-body system, so at large hyperradii, the solutions to the adiabatic eigenvalue equation correspond to the available fragmentation channels. To be more precise, we first transform the Hamiltonian into hyperspherical coordinates, and separate the hyperradial kinetic energy from the remaining Hamiltonian involving all of the angular degrees of freedom. It is simpler to understand how this works if one considers three particles moving in one dimension [12-13]. In this case, the two Jacobi coordinates, say $y_1$ and $y_2$, are simply exchanged for circular polar coordinates [12-13], and hence $R^2 = y_1^2 + y_2^2$. For four particles moving in one dimension, one simply has two hyperangles that are defined in the same way as in spherical coordinates. This amounts in general to writing the Hamiltonian as:

$$H(R,\Omega) = T(R) + H_{ad}(R,\Omega) \qquad (2)$$

The adiabatic Hamiltonian is:

$$H_{ad}(R,\Omega) = \frac{\Lambda^2}{2\mu R^2} + \sum_{i<j} V_{ij}(R,\Omega), \qquad (3)$$

where the first term is the kinetic energy in all of the angular coordinates, and the second is the sum over pair-wise interactions.

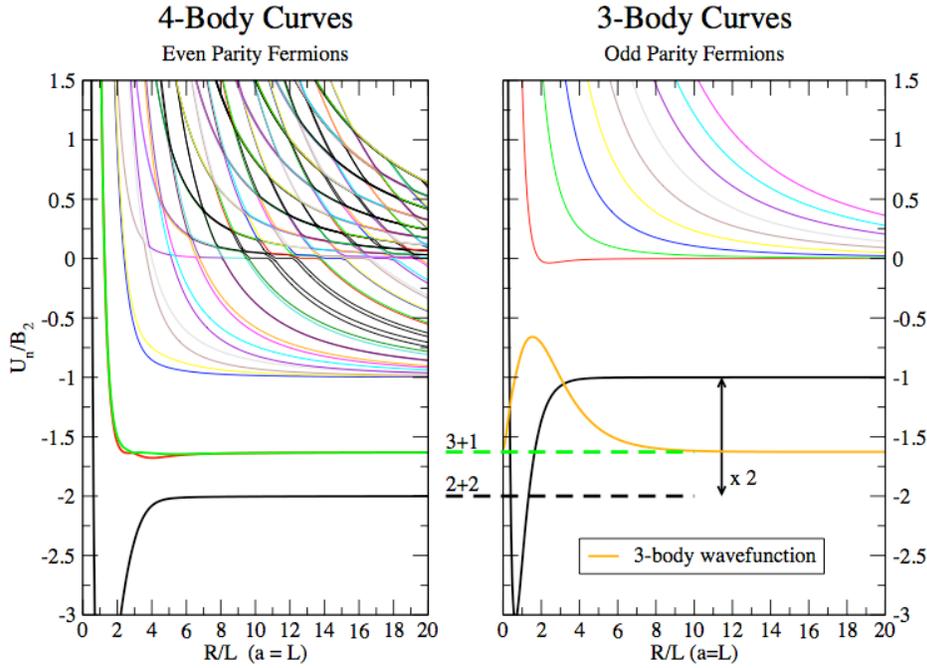

Figure 1: The adiabatic potential curves are shown for 4 even-parity fermions, and for 3 odd-parity fermions, moving in one dimension. Also shown is the ground state energy and hyperradial wavefunction for 3 odd-parity fermions.

Using a B-spline basis, it is possible to solve for the eigenstates of the adiabatic Hamiltonian for three and four particles moving in one dimension. These resulting adiabatic potential curves scaled by the two-body binding energy are shown in Fig. 1 as a function of the hyperradius scaled by the two-body even parity scattering length. Here we see clearly all of the available fragmentation channels in both the three-body and the four-body sectors. In the three-body sector, the lowest curve at large R approaches the binding energy of a single dimer, while all other curves approaching zero energy correspond to three free particles. In the four-body sector, the lowest curve at large R corresponds to two bound dimers. The next two curves correspond to a trimer and a free particle, while the series of close avoided crossings trace out the diabatic curves for four free particles. It is possible (though beyond the scope of this work) to solve these coupled adiabatic curves for cross sections governing any given elastic or inelastic process. Our goal for now is to arrive at a similar description of the four-particle system in three spatial dimensions.

## III. COORDINATE SYSTEMS AND KINEMATIC ROTATIONS

We begin by assuming that particles labeled 1 and 3 are fermions of like type, F, while particles 2 and 4 are of like type F'. There are two natural but different choices for the Jacobi coordinates of the four-particle system. We refer to them as either "H"-type Jacobi coordinates:

$$\begin{aligned}
\vec{y}_1^A &= \sqrt{\frac{\mu_{12}}{\mu}}\,(\vec{r}_1 - \vec{r}_2) \\
\vec{y}_2^A &= \sqrt{\frac{\mu_{34}}{\mu}}\,(\vec{r}_3 - \vec{r}_4) \\
\vec{y}_3^A &= \sqrt{\frac{\mu_{12,34}}{\mu}}\left(\frac{\vec{r}_1 + \vec{r}_2}{2} + \frac{\vec{r}_3 + \vec{r}_4}{2}\right)
\end{aligned} \quad (4)$$

or "K"-type Jacobi coordinates:

$$\begin{aligned}
\vec{y}_1 &= \sqrt{\frac{\mu_{12}}{\mu}}\,(\vec{r}_1 - \vec{r}_2) \\
\vec{y}_2 &= \sqrt{\frac{\mu_{12,3}}{\mu}}\left(\frac{\vec{r}_1 + \vec{r}_2}{2} - \vec{r}_3\right) \\
\vec{y}_3 &= \sqrt{\frac{\mu_{123,4}}{\mu}}\left(\frac{\vec{r}_1 + \vec{r}_2 + \vec{r}_3}{3} - \vec{r}_4\right)
\end{aligned} \quad (5)$$

Let us refer to the "H"-type coordinates defined in Eq. 4 as the "A" coordinate set. It is possible to use a different convention, "B", by simply permuting particles 2 and 3:

$$\begin{aligned}
\vec{y}_1^B &= \sqrt{\frac{\mu_{13}}{\mu}}\,(\vec{r}_1 - \vec{r}_3) \\
\vec{y}_2^B &= \sqrt{\frac{\mu_{24}}{\mu}}\,(\vec{r}_2 - \vec{r}_4) \\
\vec{y}_3^B &= \sqrt{\frac{\mu_{13,24}}{\mu}}\left(\frac{\vec{r}_1 + \vec{r}_3}{2} + \frac{\vec{r}_2 + \vec{r}_4}{2}\right)
\end{aligned} \quad (6)$$

Note that convention "A" is the more convenient choice when describing particles (1, 2) and (3, 4) in bound dimer pairs, while convention "B" is more convenient for imposing the antisymmetry between fermions of like type. Hyperspherical angles are introduced in terms of these mass-scaled Jacobi vectors as:

$$\vec{y}_i^A = (y_i^A \sin\theta_i^A \cos\phi_i^A,\, y_i^A \sin\theta_i^A \sin\phi_i^A,\, y_i^A \cos\theta_i^A) \quad (7)$$

where $y_i^A$ is the length of the $i^{\text{th}}$ mass-scaled Jacobi vector. Radial correlations are described by the hyperangles defined by:

$$\begin{aligned}
y_1^A &= R \sin\alpha_1^A \sin\alpha_2^A \\
y_2^A &= R \cos\alpha_1^A \sin\alpha_2^A \\
y_3^A &= R \cos\alpha_2^A
\end{aligned} \quad (8)$$

It is extremely useful to have at hand the transformation matrices that take us from one convention to another. For example, the mass-scaled coordinates in the "A" convention may be expressed in terms of those in the "B" convention. In order to simplify some notation, we define a matrix which has as its columns the three Jacobi vectors:

$$\mathbf{y^A} = \begin{pmatrix} y^A_{1,x} & y^A_{2,x} & y^A_{3,x} \\ y^A_{1,y} & y^A_{2,y} & y^A_{3,y} \\ y^A_{1,z} & y^A_{2,z} & y^A_{3,z} \end{pmatrix} \tag{9}$$

Then the transformation may be written as a matrix equation

$$\mathbf{y^B} = \mathbf{y^A} \mathbf{M_{AB}} \tag{10}$$

where $M_{AB}$ is an orthogonal matrix that takes us from convention "B" to "A" by expressing the "B"-coordinates in terms of the "A"-coordinates leaving the hyperradius unchanged. It is explicitly given by,

$$\mathbf{M_{AB}} = \begin{pmatrix} \frac{1}{2} & -\frac{1}{2} & \frac{1}{\sqrt{2}} \\ -\frac{1}{2} & \frac{1}{2} & \frac{1}{\sqrt{2}} \\ \frac{1}{\sqrt{2}} & \frac{1}{\sqrt{2}} & 0 \end{pmatrix}. \tag{11}$$

## IV. BODY-FIXED COORDINATES

Here we follow the reasoning of Ref. [9] and write (dropping the superscript describing the Jacobi set convention):

$$(\mathbf{y}) = (\mathbf{D}(\alpha,\beta,\gamma))^T \mathbf{\Lambda} \mathbf{D}(\Phi_1, \Phi_2, \Phi_3). \tag{12}$$

$\mathbf{D}$ is the rotation matrix:

$$\mathbf{D}(\alpha, \beta, \gamma) = \begin{pmatrix} \cos\alpha & -\sin\alpha & 0 \\ \sin\alpha & \cos\alpha & 0 \\ 0 & 0 & 1 \end{pmatrix} \begin{pmatrix} \cos\beta & 0 & \sin\beta \\ 0 & 1 & 0 \\ -\sin\beta & 0 & \cos\beta \end{pmatrix} \begin{pmatrix} \cos\gamma & -\sin\gamma & 0 \\ \sin\gamma & \cos\gamma & 0 \\ 0 & 0 & 1 \end{pmatrix} \tag{13}$$

and $\Lambda$ is a diagonal matrix whose elements are denoted $\xi_1$, $\xi_2$ and $\xi_3$. The squares of these elements are the eigenvalues of the scalar product matrix:

$$\mathbf{Y} = \mathbf{y}^T \mathbf{y}. \tag{14}$$

In democratic coordinates, they are parameterized by

$$\xi_1^2 = \frac{R^2}{3} \cos^2 \Theta_1$$
$$\xi_2^2 = \frac{R^2}{3}(3 \sin^2 \Theta_1 \sin^2 \Theta_2 + \cos^2 \Theta_1)$$
$$\xi_3^2 = \frac{R^2}{3}(3 \sin^2 \Theta_1 \cos^2 \Theta_2 + \cos^2 \Theta_1) \tag{15}$$

The factorization in Eq.12 is guaranteed to exist by singular value decomposition. The columns of the two orthogonal rotation matrices are the eigenvectors of the scalar product matrix and its transpose. The diagonal elements of $\Lambda$ are the square roots of the simultaneous eigenvalues of Y and the matrix $Z = yy^T$. The moment of inertia tensor can be expressed as:

$$\mathbf{I} = \mu(R^2\mathbf{1} - \mathbf{Y}). \tag{16}$$

The principal moments of inertia are then

$$I_1 = \frac{\mu R^2}{3}(3\sin^2\Theta_1 + 2\cos^2\Theta_1)$$
$$I_2 = \frac{\mu R^2}{3}(3\sin^2\Theta_1\cos^2\Theta_2 + 2\cos^2\Theta_1)$$
$$I_3 = \frac{\mu R^2}{3}(3\sin^2\Theta_1\sin^2\Theta_2 + 2\cos^2\Theta_1).$$

The transformation into the body-fixed frame is accomplished by the first rotation matrix in Eq. 12. In other words, the Jacobi vectors in the body-fixed frame are simply the columns of

$$\mathbf{y}' = \mathbf{\Lambda D}(\Phi_1, \Phi_2, \Phi_3) \tag{17}$$

It is necessary to write the separation distances in terms of the five internal angles. These expressions are in general "formally complicated" [9], yet they are restated here for reference (in terms of the "A" coordinate set):

$$r_{12}^2 = 2^{1/3}y_1^2$$
$$r_{13}^2 = \frac{y_1^2 + y_2^2 + 2\sqrt{2}(\vec{y_1}\cdot\vec{y_3} - \vec{y_2}\cdot\vec{y_3}) - 2\vec{y_1}\cdot\vec{y_2} + 2y_3^2}{2^{5/3}}$$
$$r_{14}^2 = \frac{y_1^2 + y_2^2 + 2\sqrt{2}(\vec{y_1}\cdot\vec{y_3} + \vec{y_2}\cdot\vec{y_3}) + 2\vec{y_1}\cdot\vec{y_2} + 2y_3^2}{2^{5/3}}$$
$$r_{23}^2 = \frac{y_1^2 + y_2^2 - 2\sqrt{2}(\vec{y_1}\cdot\vec{y_3} + \vec{y_2}\cdot\vec{y_3}) + 2\vec{y_1}\cdot\vec{y_2} + 2y_3^2}{2^{5/3}}$$
$$r_{24}^2 = \frac{y_1^2 + y_2^2 - 2\sqrt{2}(\vec{y_1}\cdot\vec{y_3} - \vec{y_2}\cdot\vec{y_3}) - 2\vec{y_1}\cdot\vec{y_2} + 2y_3^2}{2^{5/3}}$$
$$r_{34}^2 = 2^{1/3}y_2^2 \tag{18}$$

The calculation of the volume element involves the determinant of the coordinate metric g (it is actually (det g)$^{1/2}$.). For details see [9, 10]. The volume element for the eight-dimensional hypersphere is:

$$d\Omega = \frac{d\alpha\,\sin\beta\,d\beta\,d\gamma\,\sqrt{3}\cos^3\Theta_2\sin^3\Theta_2\cos 2\Theta_2\sin^9\Theta_1\sin\Phi_2}{\left[(\cos^2\Theta_1 + 3\cos^2\Theta_2\sin^2\Theta_1)(\cos^2\Theta_1 + 3\sin^2\Theta_1\sin^2\Theta_2)\right]^{1/2}} \tag{19}$$

The ranges for the angles are:

$$0 \leq \alpha \leq 2\pi$$
$$0 \leq \beta \leq \pi$$
$$0 \leq \gamma \leq 2\pi$$
$$0 \leq \Theta_1 \leq \pi$$
$$0 \leq \Theta_2 \leq \frac{\pi}{4}$$
$$0 \leq \Phi_1 \leq \pi$$
$$0 \leq \Phi_2 \leq \pi$$
$$0 \leq \Phi_3 \leq \pi \qquad (20)$$

The integration over $d\alpha \sin\beta \, d\beta \, d\gamma$ for integrands independent of these variables gives $8\pi^2$.

## V. VARIATIONAL BASIS

Current state-of-the-art three-body calculations [14-16] make use of body-fixed, so-called Smith-Whitten coordinates [17-18], which reduce the overall $J = 0$ hyperangular problem to the a two-dimensional Schrödinger equation, and then using a B-spline product basis. Typically one uses about 40-100 B-splines in each of the two dimensions, leading to 1600-10000 two-dimensional basis functions. The number of nonzero matrix elements in this basis is much smaller, about $10^5 - 10^7$, due to the fact that B-splines are piece-wise polynomials with local support. This allows for the use of banded eigenvalue solvers such as those found in the ARPACK [19] library. If this strategy were applied to the four-body problem, one would be dealing with a basis of $10^8 - 10^9$ basis functions. The banded matrix would have about $10^{11} - 10^{13}$ nonzero matrix elements! The computing power required to carry out such a calculation is presently beyond reach. It is therefore imperative to pursue a different strategy.

Our strategy is to build all of the important physics into a handful of basis functions from the beginning, allowing us to keep our basis size manageable. We choose basis functions that become exact eigenstates of the adiabatic Hamiltonian both at large and small hyperradii. At large hyperradii, the adiabatic solutions approach the available fragmentation channels of the four-body system. We have a dimer-dimer (2+2) type function:

$$\Phi_{2+2} = \mathcal{A}(\Phi_{2B}(r_{12})\Phi_{2B}(r_{34})), \qquad (21)$$

where $\mathcal{A}$ denotes the antisymmetrization operator for $S_2 \times S_2$ symmetry, and $\Phi_{2B}$ is the s-wave bound-state wavefunction for the two-particle system. We also have a dimer-three-body basis function that is written in terms of a product of a bound dimer wavefunction and a three-body hyperspherical harmonic:

$$\Phi_{2+1+1} = \mathcal{A}(\Phi_{2B}(r_{12})\mathcal{Y}_{(l_2,l_3)_{\lambda,J=0}}(\Omega)). \qquad (22)$$

At small hyperradii, the kinetic energy will dominate so that the exact eigenstates are four-body hyperspherical harmonics:

$$\Phi_{1+1+1+1} = \mathcal{A}(\mathcal{Y}_{\{(l_1,l_2)_{l_{12}},l_3\}_{\lambda,J=0}}(\Omega)). \qquad (23)$$

At intermediate hyperradii, we expect the exact eigenstates to be well described by some linear combination of these functions.

## VI. CALCULATION OF MATRIX ELEMENTS AND RESULTS

As a starting point, we consider retaining only the dimer-dimer basis function and calculating potential curve given by:

$$U_{2+2}(R) = \langle \Phi_{2+2}(R;\Omega) | H_{ad}(R,\Omega) | \Phi_{2+2}(R;\Omega) \rangle. \tag{24}$$

It is essential to include the diagonal nonadiabatic correction:

$$Q_{2+2}(R) = \langle\langle \Phi_{2+2}(R;\Omega) \left| \frac{\partial^2}{\partial R^2} \right| \Phi_{2+2}(R;\Omega) \rangle\rangle. \tag{25}$$

Here the double brackets denote integration over the angular coordinates only.

Next we address the calculation of matrix elements of the adiabatic Hamiltonian with respect to the following dimer-dimer basis function:

$$\Phi_{2+2}(\vec{r}_{12}, \vec{r}_{34}, \vec{r}_{14}, \vec{r}_{23}) = N(R)\big(\Phi_{2B}(\vec{r}_{12})\Phi_{2B}(\vec{r}_{34}) - \Phi_{2B}(\vec{r}_{23})\Phi_{2B}(\vec{r}_{14})\big) \tag{26}$$

where,

$$\Phi_{2B}(\vec{r}_{ij}) = \frac{\psi_{2B}(r_{ij})}{\sqrt{4\pi}} \tag{27}$$

and $N(R)$ is a normalization constant. Consider the kinetic energy piece:

$$\langle \Phi_{2+2} | \hat{\Lambda}^2(\Omega) | \Phi_{2+2} \rangle \tag{28}$$

We have two kinds of terms. One of the form:

$$T_1 = N(R)^2 \int d\Omega \, \frac{\psi_{2B}(r_{12})\psi_{2B}(r_{34})}{4\pi} \hat{\Lambda}^2(\Omega) \frac{\psi_{2B}(r_{12})\psi_{2B}(r_{34})}{4\pi} \tag{29}$$

and one of the form:

$$T_2 = N(R)^2 \int d\Omega \, \frac{\psi_{2B}(r_{14})\psi_{2B}(r_{23})}{4\pi} \hat{\Lambda}^2(\Omega) \frac{\psi_{2B}(r_{12})\psi_{2B}(r_{34})}{4\pi} \tag{30}$$

The total matrix element is

$$T_{2+2} = 2(T_1 - T_2) \tag{31}$$

For total J = 0, the hyperangular kinetic energy has no dependence on the three Euler angles, and the integral over dα sin β dβ dγ gives $8\pi^2$. Moreover, since we have s-wave dimers, the kinetic energy associated with the angular momentum of the three Jacobi vectors vanishes, leaving only the kinetic energy associated with radial

correlations [20]:

$$\hat{\Lambda}^2 \psi_{2B}(r_{12})\psi_{2B}(r_{34}) = \\
\frac{R}{2^{5/6}}\left(\frac{(1-5\cos 2\alpha_1)}{\sin\alpha_2 \sin\alpha_1} - \frac{2\sin\alpha_1(1+4\cos 2\alpha_2)}{\sin\alpha_2}\right)\psi'_{2B}(r_{12})\psi_{2B}(r_{34}) \\
+ \frac{R}{2^{5/6}}\left(\frac{(1+5\cos 2\alpha_1)}{\sin\alpha_2 \cos\alpha_1} - \frac{2\cos\alpha_1(1+4\cos 2\alpha_2)}{\sin\alpha_2}\right)\psi_{2B}(r_{12})\psi'_{2B}(r_{34}) \\
+ 2^{1/3}R^2 \sin 2\alpha_1 \sin^2\alpha_2 \psi'_{2B}(r_{12})\psi'_{2B}(r_{34}) \\
- 2^{1/3}R^2 \left(\cos^2\alpha_1 + \cos^2\alpha_2 \sin^2\alpha_1\right)\psi''_{2B}(r_{12})\psi_{2B}(r_{34}) \\
- 2^{1/3}R^2 \left(\cos^2\alpha_1 \cos^2\alpha_2 + \sin^2\alpha_1\right)\psi_{2B}(r_{12})\psi''_{2B}(r_{34})$$
(32)

The integration must be carried out in the five democratic angles, so it is necessary to determine the hyperangles $\alpha_1$ and $\alpha_2$ given the angles $\Phi_1, \Phi_2, \Phi_3, \Theta_1, \Theta_2$ as input. This is readily accomplished by first using Eq. 17 in order to determine the body-fixed Jacobi-vectors, then using Eq. 8.

In the top-left graph appearing in Fig. 2, we show the total hyperradial potential felt by two interacting dimers,

$$W_{2+2}(R) = U_{2+2}(R) - Q_{2+2}(R)/(2\mu).$$
(33)

for the case where the atom-atom scattering length $a = 10L$. Note that the inclusion of the nonadiabatic correction is absolutely essential to give the correct large $R$ behavior. To be more precise, the $1/R^2$ coefficient in the potential $U_{2+2}$ at large $R$ is exactly canceled by the $1/R^2$ coefficient in $Q_{2+2}/(2\mu)$ making it possible to calculate an s-wave scattering phase-shift. The graph appearing on the right side of Fig. 2 shows the quantity $(k_{2+2}a)\cot(\delta)$ ($k_{2+2} = (2\mu_{2+2}E)^{1/2}$ is the dimer-dimer momentum wavenumber) for elastic dimer-dimer scattering. In the limit $a \gg L$, we expect $a_{DD} = 0.6a$ [8]. We find for $a = 10L$, where $L$ is the range of the potential in Eq. 1, that $a_{DD} \approx 0.8a$, in remarkably good agreement with [8] for a variational calculation with only one basis function. We are currently conducting calculations for larger $a$ in an attempt to see a universal trend towards the expected [8] result $a_{DD} = 0.6a$, but these calculations are technically difficult due to the computational demands of the five-dimensional integrals. Our calculations, however, do have the advantage that they are valid away from the universal regime, and that they readily give the energy dependence of the dimer-dimer scattering length as shown in right graph appearing in Fig. 2.

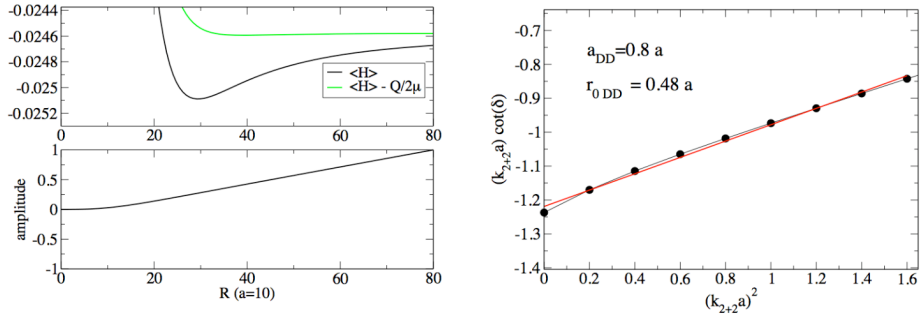

Figure 2: Top-left: the lowest adiabatic potential curve corresponding to elastic dimer-dimer scattering without the diagonal correction (black), and with the diagonal correction (green). Bottom-left: the wavefunction for elastic dimer-dimer scattering at threshold. Right: the energy dependence of the quantity $(ka)cot(\delta)$ is shown. The connected black points are the present calculations, while the red line is a linear fit giving our preliminary values for the scattering length and effective range. We've used $L=m=1$ in all of these calculations, and $a=10L$ for the data shown here.

We have performed a more sophisticated variational calculation involving five basis functions: one (2+2) function, three (2+1+1) functions and one (1+1+1+1) function. We have retained only three-body harmonics in the product 2+1+1 functions and 1+1+1+1 four-body harmonics with hyperangular momentum quantum number $\lambda \leq 2$. The results of this calculation for $a = 2$ and $a = 10$ are shown in Fig. 4. The dashed lines are simply diagonal expectation values of the adiabatic Hamiltonian, normalized by the overlap integral. The solid lines are the diagonalized eigenenergies in this 5-function basis set. There are several things to note here. First, the diagonalized ground state eigenenergy is always lower than the expectation value, as we expect from the variational principle. Second, the addition of four extra basis functions has little effect on the ground state potential curve, somewhat justifying our single channel dimer-dimer calculations in Fig. 2. It is also clear from the three-body curves, that there is no three-body bound state, even for $a = 2$. This justifies the exclusion of a basis function involving a trimer and a free particle in our variational four-body calculation.

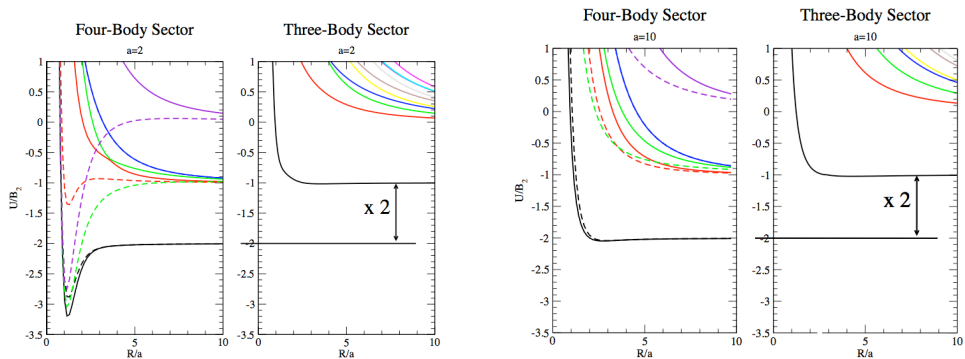

Figure 3: The adiabatic potential curves for identical fermions in two spin substates, moving in three dimensions are shown. All calculations have been performed in units such that *L=m=1*. Left: Four-body and three-body curves for *a=2L*. Right: Four-body and three-body curves for *a=10L*.

VII. CONCLUSIONS

In conclusion, we have performed a variational calculation of the adiabatic hyperspherical potential curves for a two-component four-fermion system. Our approach makes use of basis functions that become exact eigenstates of the adiabatic Hamiltonian both at very small and very large hyperradii. We have incorporated all of the important physics for the low energy range considered here into a handful of functions, thereby allowing us to carry out calculations with only 5 by 5 matrices.

Looking forward, there remains a great deal to do. We must calculate the first-derivative coupling matrix in order to complete the hyperspherical picture. These couplings will then allow us to calculate reaction rates for inelastic processes such as (1+1+1+1 → 2+1+1), (2+1+1 → 2+2) or the reverse. We are presently performing calculations for larger scattering lengths in order to make a more quantitative comparison with the results of Ref. [8]. Finally, in order to fully understand the physics of the BEC/BCS crossover, we must explore the energy landscape on the negative scattering length side of the Feshbach resonance. This will involve a modification of our variational basis to include a function that correctly represents atom-atom scattering near threshold.

**Acknowledgements**

The authors gratefully acknowledge Brett Esry for helpful discussions, and the National Science Foundation for partial support.

**References**

[1] C. A. Regal, C. Ticknor, J. L. Bohn, and D. S. Jin. *Nature*, **424**, 47–50 (2003).
[2] K. E. Strecker, G. B. Partridge, and R. G. Hulet. *Phys. Rev. Lett.,* **91**, 80406 (2003).
[3] M. Greiner, C. A. Regal, and D. S. Jin. *Nature*, **426**, 537–540 (2003).
[4] S. Jochim, M. Bartenstein, A. Altmeyer, G. Hendl, S. Riedl, C. Chin, J. Hecker Denschlag, and R. Grimm. *Science*, **302**, 2101–2104 (2003).
[5] M. W Zwierlein, CA Stan, CH Schunck, SMF Raupach, S. Gupta, Z. Hadzibabic, and W. Ketterle. *Phys. Rev. Lett.,* **91**, 250401 (2003).
[6] M. J. Holland, S. Kokkelmans, ML Chiofalo, and R. Walser. *Phys. Rev. Lett.,* **87**, 120406 (2001).
[7] M. J. Holland, C. Menotti, and L. Viverit. Arxiv preprint cond-mat/0404234 (2004).
[8] D. S. Petrov, C. Salomon, and G. V. Shlyapnikov. *Phys. Rev. Lett.,* **93**, 90404 (2004).
[9] V. Aquilanti and S. Cavalli. *J. Chem. Soc. : Faraday Transactions*, **95**, 801-809 (1997).
[10] A. Kuppermann. *J. Phys. Chem. A*, **101**, 6368 (1997).
[11] J. Macek. *J. Phys. B*, **1**,831–843 (1968).
[12] W. Gibson, S. Y. Larsen, and J. Popiel. Phys. Rev. A, **35**, 4919 (1987).
[13] N. P. Mehta and J. R. Shepard. *Phys. Rev. A*, **72**, 032728, (2005). N. P. Mehta, B. D. Esry and C. H. Greene, (unpublished).
[14] J. P. D'Incao and B. D. Esry. *Phys. Rev. Lett.,* **94**, 213201 (2005).
[15] B. D. Esry, C. H. Greene, and J. P. Burke Jr. *Phys. Rev. Lett.,* **83**, 1751–1754 (1999).
[16] H. Suno, B. D Esry, C.H. Greene, and J.P. Burke Jr. *Phys. Rev. A*, **65**, 42725, (2002).

[17] F. T. Smith, *J. Math Phys*. **3**, 735 (1962).
[18] R. C. Whitten and F. T. Smith *J. Math. Phys*. **9** 1103 (1968).
[19] R. B. Lehoucq, D. C. Sorensen, and C. Yang. ARPACK Users' Guide: Solution of Large-scale Eigenvalue Problems with Implicitly Restarted Arnoldi Methods. SIAM (1998).
[20] C. H. Greene and C. W. Clark. *Phys. Rev. A*, **30**, 2161 (1984).